# Load Balancing for MapReduce-based Entity Resolution

Lars Kolb, Andreas Thor, Erhard Rahm

*Database Group, University of Leipzig*
*Leipzig, Germany*
{kolb,thor,rahm}@informatik.uni-leipzig.de

*Abstract*—The effectiveness and scalability of MapReduce-based implementations of complex data-intensive tasks depend on an even redistribution of data between map and reduce tasks. In the presence of skewed data, sophisticated redistribution approaches thus become necessary to achieve load balancing among all reduce tasks to be executed in parallel. For the complex problem of entity resolution, we propose and evaluate two approaches for such skew handling and load balancing. The approaches support blocking techniques to reduce the search space of entity resolution, utilize a preprocessing MapReduce job to analyze the data distribution, and distribute the entities of large blocks among multiple reduce tasks. The evaluation on a real cloud infrastructure shows the value and effectiveness of the proposed load balancing approaches.

## I. INTRODUCTION

Cloud computing [2] has become a popular paradigm for efficiently processing computationally and data-intensive tasks. Such tasks can be executed on demand on powerful distributed hardware and service infrastructures. The parallel execution of complex tasks is facilitated by different programming models, in particular the widely available MapReduce (MR) model [5] supporting the largely transparent use of cloud infrastructures. However, the (cost-) effectiveness and scalability of MR implementations depend on effective load balancing approaches to evenly utilize available nodes. This is particularly challenging for data-intensive tasks where skewed data redistribution may cause node-specific bottlenecks and load imbalances.

We study the problem of MR-based load balancing for the complex problem of entity resolution (ER) (also known as object matching, deduplication, record linkage, or reference reconciliation), i.e., the task of identifying entities referring to the same real-world object [13]. ER is a pervasive problem and of critical importance for data quality and data integration, e.g., to find duplicate customers in enterprise databases or to match product offers for price comparison portals. ER techniques usually compare pairs of entities by evaluating multiple similarity measures to make effective match decisions. Naïve approaches examine the complete Cartesian product of $n$ input entities. However, the resulting quadratic complexity of $O(n^2)$ is inefficient for large datasets even on cloud infrastructures. The common approach to improve efficiency is to reduce the search space by adopting so-called blocking techniques [3]. They utilize a blocking key on the values of one or several entity attributes to partition the input data into multiple partitions (called blocks) and restrict the subsequent matching to entities of the same block. For example, product entities may be partitioned by manufacturer values such that only products of the same manufacturer are evaluated to find matching entity pairs.

Despite the use of blocking, ER remains a costly process that can take several hours or even days for large datasets [12]. Entity resolution is thus an ideal problem to be solved in parallel on cloud infrastructures. The MR model is well suited to execute blocking-based ER in parallel within several map and reduce tasks. In particular, several map tasks can read the input entities in parallel and redistribute them among several reduce tasks based on the blocking key. This guarantees that all entities of the same block are assigned to the same reduce task so that different blocks can be matched in parallel by multiple reduce tasks.

However, such a basic MR implementation is susceptible to severe load imbalances due to skewed blocks sizes since the match work of entire blocks is assigned to a single reduce task. As a consequence, large blocks (e.g., containing 20% of all entities) would prevent the utilization of more than a few nodes. The absence of skew handling mechanisms can therefore tremendously deteriorate runtime efficiency and scalability of MR programs. Furthermore, idle but instantiated nodes may produce unnecessary costs because public cloud infrastructures (e.g., Amazon EC2) usually charge per utilized machine hours.

In this paper, we propose and evaluate two effective load balancing approaches to data skew handling for MR-based entity resolution. Note that MR's inherent vulnerability to load imbalances due to data skew is relevant for all kind of pairwise similarity computation, e.g., document similarity computation [9] and set-similarity joins [19]. Such applications can therefore also benefit from our load balancing approaches though we study MR-based load balancing in the context of ER only. In particular, we make the following contributions:

- We introduce a general MR workflow for load-balanced blocking and entity resolution. It employs a preprocessing MR job to determine a so-called block distribution matrix that holds the number of entities per block separated by input partitions. The matrix is used by both load balancing schemes to determine fine-tuned entity redistribution for parallel matching of blocks. (Section III)
- The first load balancing approach, BlockSplit, takes the size of blocks into account and assigns entire blocks to

reduce tasks if this does not violate load balancing or memory constraints. Larger blocks are split into smaller chunks based on the input partitions to enable their parallel matching within multiple reduce tasks. (Section IV)
- The second load balancing approach, PairRange, adopts an enumeration scheme for all pairs of entities to evaluate. It redistributes the entities such that each reduce task has to compute about the same number of entity comparisons. (Section V)
- We evaluate our strategies and thereby demonstrate the importance of skew handling for MR-based ER. The evaluation is done on a real cloud environment, uses real-world data, and compares the new approaches with each other and the basic MR strategy. (Section VI)

In the next section we review the general MR program execution model. Related work is presented in Section VII before we conclude. Furthermore, we describe an extension of our strategies for matching two sources in Appendix I. Appendix II lists the pseudo-code for all proposed algorithms.

## II. MAPREDUCE PROGRAM EXECUTION

MapReduce (MR) is a programming model designed for parallel data-intensive computing in cluster environments with up to thousands of nodes [5]. Data is represented by key-value pairs and a computation is expressed with two user defined functions:

$$\text{map} : (key_{in}, value_{in}) \rightarrow list(key_{tmp}, value_{tmp})$$
$$\text{reduce} : (key_{tmp}, list(value_{tmp})) \rightarrow list(key_{out}, value_{out})$$

These functions contain sequential code and can be executed in parallel on disjoint partitions of the input data. The map function is called for each input key-value pair whereas reduce is called for each key $key_{tmp}$ that occurs as map output. Within the reduce function one can access the list of all corresponding values $list(value_{tmp})$.

Besides map and reduce, a MR dataflow relies on three further functions. First, the function part partitions the map output and thereby distributes it to the available reduce tasks. All keys are then sorted with the help of a comparison function comp. Finally, each reduce task employs a grouping function group to determine the data chunks for each reduce function call. Note that each of these functions only operates on the key of key-value pairs and does not take the values into account. Keys can have an arbitrary structure and data type but need to be comparable. The use of extended (composite) keys and an appropriate choice of part, comp, and group supports sophisticated partitioning and grouping behavior and will be utilized in our load balancing approaches.

For example, the center of Figure 1 shows an example MR program with two map tasks and three reduce tasks. The map function is called for each of the four input key-value pairs (denoted as ■) and the map phase emits an overall of 10 key-value pairs using composite keys (Figure 1 only shows keys for simplicity). Each composite key has a shape (circle or triangle) and a color (light-gray, dark-gray, or black). Keys are assigned to three reduce tasks using a partition function that is only based on a part of the key ("color"). Finally, the group function employs the entire key so that the reduce function is called for 5 distinct keys.

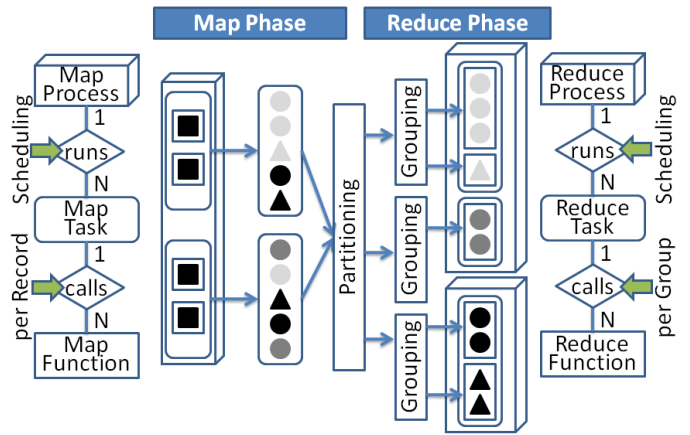

Fig. 1. Schematic overview of example MR program execution using 1 map process, $m$=2 map tasks, 2 reduce processes, and $r$=3 reduce tasks. In this example, partitioning is based on the key's color only and grouping is done on the entire key.

The actual execution of an MR program (also known as job) is realized by an MR framework implementation such as Hadoop [1]. An MR cluster consists of a set of nodes that run a fixed number of map and reduce processes. For each MR job execution, the number of map tasks ($m$) and reduce tasks ($r$) is specified. Note that the partition function part relies on the number of reduce tasks since it assigns key-value pairs to the available reduce tasks. Each process can execute only one task at a time. After a task has finished, another task is automatically assigned to the released process using a framework-specific scheduling mechanism. The example MR program of Figure 1 runs in a cluster with one map and two reduce processes, i.e., one map task and two reduce tasks can be processed simultaneously. Hence, the only map process runs two map tasks and the three reduce tasks are eventually assigned to two reduce processes.

## III. LOAD BALANCING FOR ER

We describe our load balancing approaches for ER for one data source $R$. The input is a set of entities and the output is a match result, i.e., pairs of entities that are considered to be the same. With respect to blocking, we assume that all entities have a valid blocking key. The generalization to consider entities without defined blocking key (e.g. missing manufacturer information for products) is relatively easy. All entities $R_\emptyset \subseteq R$ without blocking key need to be matched with all entities, i.e., the Cartesian product of $R \times R_\emptyset$ needs to be determined which is a special case of ER between two sources. The Appendix explains how our strategies can be extended for matching two sources.

As discussed in the introduction, parallel ER using blocking can be easily implemented with MR. The map function can be used to determine for every input entity its blocking key and to output a key-value pair (blocking_key, entity). The default partitioning strategy would use the blocking key to distribute key-value pairs among reduce tasks so that all entities sharing the same blocking key are assigned to the same reduce task. Finally, the reduce function is called for each block and computes the matching entity pairs within its block. We call this straightforward approach Basic. However, the Basic strategy is vulnerable to data skew due to blocks of largely varying size. Therefore the execution time may be dominated by a single or a few reduce tasks. Processing large blocks may also lead to serious memory problems because entity resolution requires that all entities within the same block are compared with each other. A reduce task must therefore store all entities passed to a reduce call in main memory – or must make use of external memory which further deteriorates execution times.

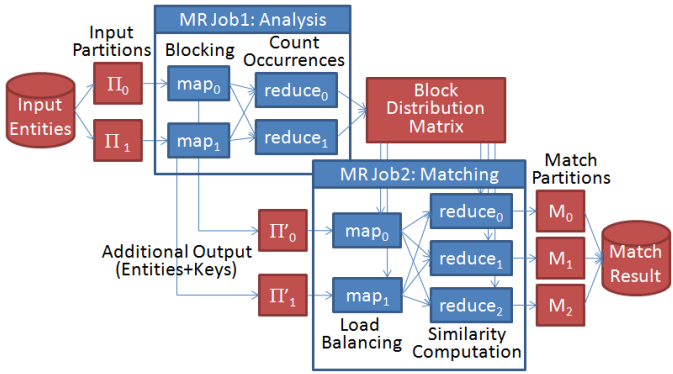

Fig. 2. Schematic overview of the MR-based matching process with load balancing.

| Partition | $\Pi_0$ | | | | | | | $\Pi_1$ | | | | | | |
|---|---|---|---|---|---|---|---|---|---|---|---|---|---|---|
| Entity | A | B | C | D | E | F | G | H | I | K | L | M | N | O |
| Blocking Key | w | w | x | x | x | z | z | w | w | y | y | z | z | z |

Fig. 3. The example data consists of 14 entities $A$-$O$ that are divided into two partitions $\Pi_0$ and $\Pi_1$.

A domain expert might, of course, adjust the blocking function so that it returns blocks of similar sizes. However, this tuning is very difficult because it must ensure that matching entities still reside in the same block. Furthermore, the blocking function needs to be adjusted for every match problem individually. We therefore propose two general load balancing approaches that address the mentioned skew and memory problems by distributing the processing of large blocks among several reduce tasks. Both approaches are based on a general ER workflow with two MR jobs that is described next. The first MR job, described in Section III-B, analyzes the input data and is the same for both load balancing schemes. The different load balancing strategies BlockSplit and PairRange are described in the following sections IV and V, respectively.

### A. General ER Workflow for Load Balancing

To realize our load balancing strategies, we perform ER processing within two MR jobs as illustrated in Figure 2. Both jobs are based on the same number of map tasks and the same partitioning of the input data.[1] The first job calculates a so-called block distribution matrix (BDM) that specifies the number of entities per block separated by input partitions. The matrix is used by the load balancing strategies (in the second

[1] See Appendix II for details.

MR job) to tailor entity redistribution for parallel matching of blocks of different size.

Load balancing is mainly realized within the map phase of the second MR job. Both strategies follow the idea that map generates a carefully constructed composite key that (together with associated partition and group functions) allows a balanced load distribution. The composite key thereby combines information about the target reduce task(s), the block of the entity, and the entity itself. While the MR partitioning may only use part of the map output key for routing, it still groups together key-value pairs with the same blocking key component of the composite key and, thus, makes sure that only entities of the same block are compared within the reduce phase. As we will see, the map function may generate multiple keys per entity if this entity is supposed to be processed by multiple reduce tasks for load balancing. Finally, the reduce phase performs the actual ER and computes match similarities between entities of the same block. Since the reduce phase consumes the vast majority of the overall runtime (more than 95% in our experiments), our load balancing strategies solely focus on data redistribution for reduce tasks. Other MR-specific performance factors are therefore not considered. For example, consideration of data locality (see, e.g., [10]) would have only limited impact and would require additional modification of the MR framework.

### B. Block Distribution Matrix

The block distribution matrix (BDM) is a $b \times m$ matrix that specifies the number of entities of $b$ blocks across $m$ input partitions. The BDM computation using MR is straightforward. The map function determines the blocking key for each entity and outputs a key-value pair with a composite map output key (blocking_key $\odot$ partition_index) and a corresponding value of 1 for each entity[2]. The key-value pairs are partitioned based on the blocking key component to ensure that all data for a specific block is processed in the same reduce task. The reduce task's key-value pairs are sorted and grouped by the entire key and reduce counts the number of blocking keys (i.e., entities per block) per partition and outputs triples of the form (blocking key, partition index, number of entities).

For illustration purposes, we use a running example with 14 entities and 4 blocking keys as shown in Figure 3. Figure 4 illustrates the computation of the BDM for this example data. So the map output key of $M$ is $z.1$ because $M$'s blocking key equals $z$ and $M$ appears in the second partition (partition index=1). This key is assigned to the last reduce task that outputs $[z, 1, 3]$ because there are 3 entities in the second partition for blocking key $z$. The combined reduce outputs correspond to a row-wise enumeration of non-zero matrix cells. To assign block keys to rows of the BDM, we use the (arbitrary) order of the blocks from the reduce output, i.e., we assign the first block (key $w$) to block index position 0, etc. The block sizes in the example vary between 2 and 5 entities. The match work to compare all entities per block with each other thus ranges from 1 to 10 pair comparisons; the largest block with key $z$ entails 50% of all comparisons although it contains only 35% (5 of 14) of all entities.

[2] A combine function that aggregates the frequencies of the blocking keys per map task might be employed as an optimization.

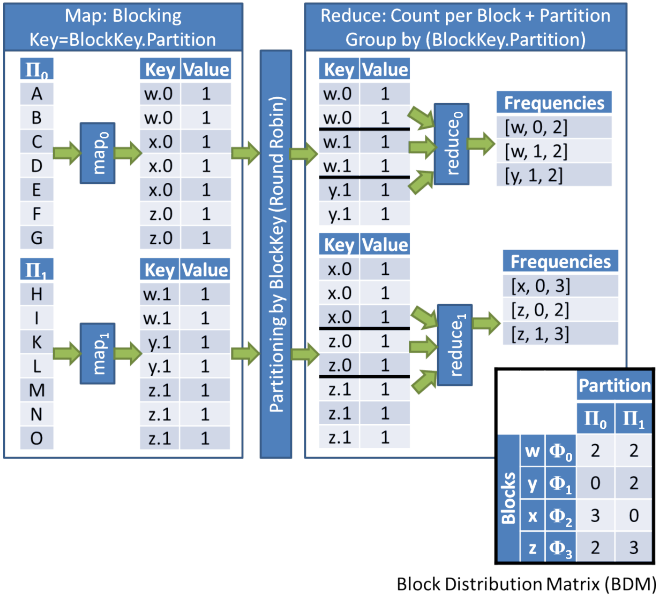

Fig. 4. Example dataflow for computation of the block distribution matrix (MR Job1 of Figure 2) using the example data of Figure 3.

As illustrated in Figure 2, map produces an additional output $\Pi_i'$ per partition that contains the original entities annotated with their blocking keys. This output is not shown in Figure 4 to save space but used as input in the second MR job (see Figures 5 and 7).

## IV. Block-based Load Balancing

The first strategy, BlockSplit, generates one or several so-called match tasks per block and distributes match tasks among reduce tasks. Furthermore, it uses the following two ideas:

- BlockSplit processes small blocks within a single match tasks similar to the basic MR implementation. Large blocks are split according to the $m$ input partitions into $m$ sub-blocks. The resulting sub-blocks are then processed using match tasks of two types. Each sub-block is (like any unsplit block) processed by a single match task. Furthermore, pairs of sub-blocks are processed by match tasks that evaluate the Cartesian product of two sub-blocks. This ensures that all comparisons of the original block will be computed in the reduce phase.
- BlockSplit determines the number of comparisons per match task and assigns match tasks in descending size among reduce tasks. This implements a greedy load balancing heuristic ensuring that the largest match tasks are processed first to make it unlikely that they dominate or increase the overall execution time.

The realization of BlockSplit makes use of the BDM as well as of composite map output keys. The map phase outputs key-value pairs with key=(reduce_index ⊙ block_index ⊙ split) and value=(entity). The reduce task index is a value between 0 and $r-1$ is used by the partition function to realize the desired assignment to reduce tasks. The grouping is done on the entire key and – since the block index is part of the key – ensures that each reduce function only receives entities of the same block. The split value indicates what match task has to be performed by the reduce function, i.e., whether a complete block or sub-blocks need to be processed. In the following, we describe map key generation in detail.

During the initialization, each of the $m$ map tasks reads the BDM and computes the number of comparison per block and the total number of comparisons $P$ over all $b$ blocks $\Phi_k$: $P = \frac{1}{2} \cdot \Sigma_{k=0}^{b-1} |\Phi_k| \cdot (|\Phi_k|-1)$. For each block $\Phi_k$, it also checks if the number of comparisons is above the average reduce task workload, i.e., if

$$\frac{1}{2} \cdot |\Phi_k| \cdot (|\Phi_k| - 1) > P/r$$

If the block $\Phi_k$ is *not* above the average workload it can be processed within a single match task (this is denoted as $k.*$ in the block_index and split components of the map output key). Otherwise it is split into $m$ sub-blocks based on the $m$ input partitions[3] leading to the following $\frac{1}{2} \cdot m \cdot (m-1) + m$ match tasks:

- $m$ match tasks, denoted with key components $k.i$, for the individual processing of the $i^{\text{th}}$ sub-block for $i \in [0, m-1]$
- $\frac{1}{2} \cdot m \cdot (m-1)$ match tasks, denoted with key components $k.i \times j$ with $i, j \in [0, m-1]$ and $i < j$, for the computation of the Cartesian product of sub-blocks $i$ and $j$

To determine the reduce task for each match task, all match tasks are first sorted in descending order of their number of comparisons. Match tasks are then assigned to reduce tasks in this order so that the current match task is assigned to the reduce task with the lowest number of already assigned pairs. In the following, we denote the reduce task index for match task $k.x$ with $R(k.x)$.

After the described initialization phase, the map function is called for each input entity. If the entity belongs to a block $\Phi_k$ that has *not* to be split, map outputs one key-value pair with composite key=$R(k.*).k.*$. Otherwise, map outputs $m$ key-value pairs for the entity. The key $R(k.i).k.i$ represents the individual sub-block $i$ of block $\Phi_k$ and the remaining $m-1$ pairs represent all combinations with the other $m-1$ sub-blocks. This indicates that entities of split blocks are replicated $m$ times to support load balancing. The map function emits the entity as value of the key-value pair; for split blocks we annotate entities with the partition index for use in the reduce phase.

---

[3]Note that the BDM holds the number of entities per (block, partition) pair and map can therefore determine which input partitions contain entities of $\Phi_k$. However, in favor of readability we assume that all $m$ input partitions contain at least one entity. Our implementation, of course, ignores unnecessary partitions.

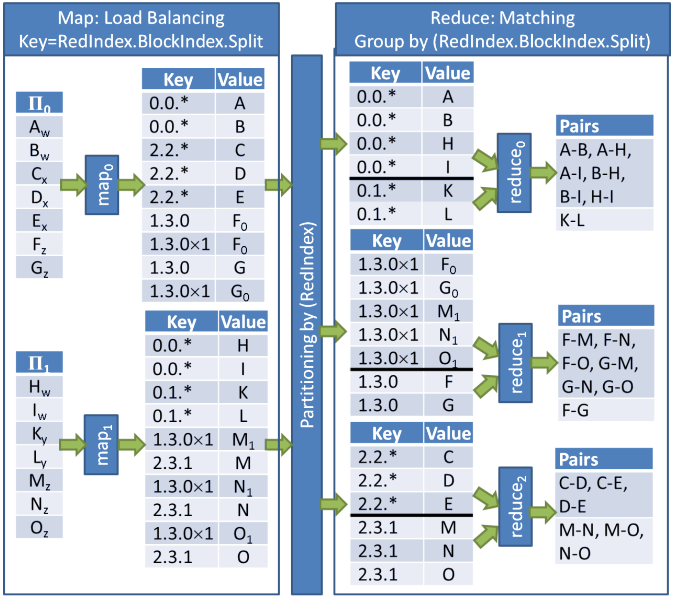

Fig. 5. Example dataflow for the load balancing strategy BlockSplit.

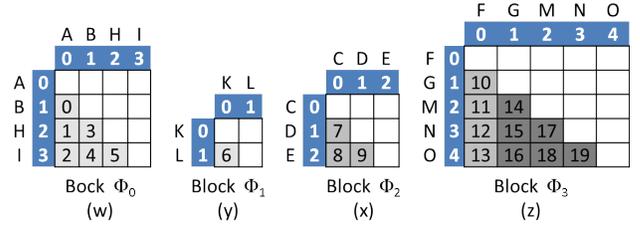

Fig. 6. Global enumeration of all pairs for the running example. The three different shades indicate how the PairRange strategy assigns pairs to 3 reduce tasks.

In our running example, only block $\Phi_3$ (blocking key $z$) is subject to splitting into $m=2$ sub-blocks. The BDM (see Figure 4) indicates for block $\Phi_3$ that $\Pi_0$ and $\Pi_1$ contain 2 and 3 entities, respectively. The resulting sub-blocks $\Phi_{3.0}$ and $\Phi_{3.1}$ lead to the three match tasks *3.0*, *3.0×1*, and *3.1* that account for 1, 6, and 3 comparisons, respectively. The resulting ordering of match tasks by size (*0.\**, *3.0×1*, *2.\**, *3.1*, *1.\**, and *3.0*) leads for three reduce tasks to the distribution shown in Figure 5. The replication of the five entities for the split block leads to 19 key-value pairs for the 14 input entities. Each reduce task has to process between six and seven comparisons indicating a good load balancing for the example.

## V. PAIR-BASED LOAD BALANCING

The block-based strategy BlockSplit splits large blocks according to the input partitions. This approach may still lead to unbalanced reduce task workloads due to differently-sized sub-blocks. We therefore propose a more sophisticated pair-based load balancing strategy PairRange that targets at a uniform number of pairs for all reduce tasks. It uses the following two ideas:

- PairRange implements a virtual enumeration of all entities and relevant comparisons (pairs) based on the BDM. The enumeration scheme is used to sent entities to one or more reduce tasks and to define the pairs that are processed by each reduce tasks.
- For load balancing, PairRange splits the range of all relevant pairs into $r$ (almost) equally sized pair ranges and assigns the $k^{\text{th}}$ range $\Re_k$ to the $k^{\text{th}}$ reduce task.

Each map task processes its input partition row-by-row and can therefore enumerate entities per partition and block. Although entities are processed independently in different partitions, the BDM permits to compute the global block-specific entity index locally within the map phase. Given a partition $\Pi_i$ and a block $\Phi_k$, the overall number of entities of $\Phi_k$ in all preceding partitions $\Pi_0$ through $\Pi_{i-1}$ has just to be added as offset. For example, entity $M$ is the first entity of block $\Phi_3$ in partition $\Pi_1$. Since the BDM indicates that there are two other entities in $\Phi_3$ in the preceding partition $\Pi_0$, $M$ is the third entity of $\Phi_3$ and is thus assigned entity index 2. Figure 6 shows block-wise the resulting index values for all entities of the running example (white numbers).

Enumeration of entities allows for an effective enumeration of all pairs to compare. An entity pair $(x,y)$ with entity indexes $x$ and $y$ is only enumerated if $x < y$. We thereby avoid unnecessary computation, i.e., pairs of the same entity $(x,x)$ are not considered as well as pairs $(y,x)$ if $(x,y)$ has already been considered. Pair enumeration employs a column-wise continuous enumeration across all blocks based on information of the BDM. The pair index $p_i(x,y)$ of two entities with indexes $x$ and $y$ ($x < y$) in block $\Phi_i$ is defined as follows:

$$p_i(x,y) = c(x,y,|\Phi_i|) + o(i) \quad (1)$$

with $c(x,y,N) = \frac{x}{2}(2 \cdot N - x - 3) + y - 1$ and $o(i) = \frac{1}{2} \cdot \sum_{k=0}^{i-1}(|\Phi_k| \cdot (|\Phi_k| - 1))$. Here $c(x,y,N)$ is the index of the cell $(x,y)$ in an $N \times N$ matrix and $o(i)$ is the offset and equals the overall number of pairs in all preceding blocks $\Phi_0$ through $\Phi_{i-1}$. The number of entities in block $\Phi_i$ is denoted as $|\Phi_i|$. Figure 6 illustrates the pair enumeration for the running example. The pair index of pair $p_i(x,y)$ can be found in the column $x$ and row $y$ of block $\Phi_i$. For example, the index for pair $(2,3)$ of block $\Phi_0$ equals 5.

PairRange splits the range of all pairs into $r$ almost equally-sized pair ranges and assigns the $k^{\text{th}}$ range $\Re_k$ to the $k^{\text{th}}$ reduce task. $k$ is therefore both, the reduce task index and the range index. Given a total of $P$ pairs and $r$ ranges, a pair with index $0 \leq p < P$ falls in $\Re_k$ if

$$p \in \Re_k \iff k = \lfloor r \cdot \frac{p}{P} \rfloor \quad (2)$$

The first $r-1$ reduce tasks processes $\lceil \frac{P}{r} \rceil$ pairs each whereas the last reduce task is responsible for the remaining $P - (r-1) \cdot \lceil \frac{P}{r} \rceil$ pairs. In the example of Figure 6, we have $P = 20$ pairs, so that for $r = 3$ we obtain the ranges reduce task's pair range, and – if this is the case – computes the matching for this pair. To this end, map additionally annotates each entity with its entity index so that the pair index can be easily computed by the reduce function.

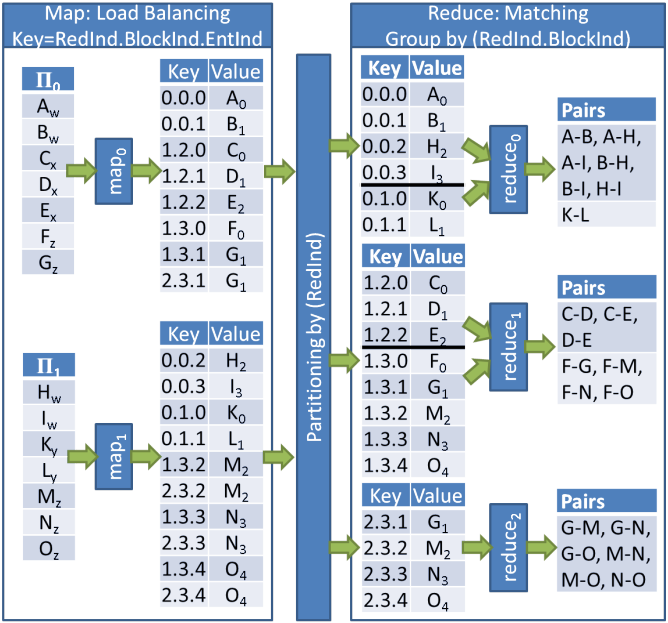

Fig. 7. Example dataflow for the load balancing strategy PairRange.

Figure 7 illustrates the PairRange strategy for our running example. Entity $M$ belongs to block $\Phi_3$, has an entity index of 2, and takes part in 4 pairs with pair indexes 11, 14, 17, and 18, respectively. Given the three ranges $[0, 6]$, $[7, 13]$, and $[14, 19]$, entity $M$ has to be sent to the second reduce task (index=1) for pair #11 and the third reduce task (index=2) for the other pairs. map therefore outputs two tuples $(1.3.2, M)$ and $(2.3.2, M)$. The second reduce task not only receives $M$ but all entities of $\Phi_3$ ($F$, $G$, $M$, $N$, and $O$). However, due to its assigned pair range $[7, 13]$, it only processes pairs with indexes 10 through 13 of $\Phi_3$ (and, of course, 7 through 9 of $\Phi_2$). The remaining pairs of $\Phi_3$ are processed by the third reduce task which receives all entities of $\Phi_3$ but $F$ because the latter does not take part in any of the pairs with index 14 through 19 (see Figure 6).

$\Re_0 = [0, 6]$, $\Re_1 = [7, 13]$, and $\Re_2 = [14, 19]$ (illustrated by different shades).

During the initialization, each of the $m$ map tasks reads the BDM, computes the total number of comparisons $P$, and determines the $r$ pair ranges. Afterwards, the map function is called for each entity $e$ and determines $e$'s entity index $x$ as well as all relevant ranges, i.e., all ranges that contain at least one pair where $e$ is participating. The identification of relevant ranges does *not* require the examination of the possibly large number of all pairs but can be mostly realized by processing two pairs. Let $N$ be the size of $e$'s block, entity $e$ with index $x$ takes part in the pairs $(0, x), \ldots, (x-1, x), (x, x+1), \ldots, (x, N-1)$. The enumeration scheme thus allows for a quick identification of $p_{min}$ and $p_{max}$, i.e., $e$'s pairs with the smallest and highest pair index. For example, $M$ has an entity index of 2 within a block of size $|\Phi_3| = 5$ and the two pairs are therefore $p_{min} = p_3(0, 2) = 11$ and $p_{max} = p_3(2, 4) = 18$. All relevant ranges of $e$ are between $\Re_{min} \ni p_{min}$ and $\Re_{max} \ni p_{max}$ because the range index is monotonically increasing with the pair index (see formula (2)). Entity $M$ is thus only needed for the second and third pair range (reduce task).

Finally, map emits a key-value pair with key= (range_index $\odot$ block_index $\odot$ entity_index) and value=entity for each relevant range. The MR partitioning is based on the range index only for routing all data of range $\Re_k$ to the reduce task with index $k$. The sorting is done based on the entire key whereas the grouping is done by range index and block index. The reduce task does not necessarily receive all entities of a block but only those entities that are relevant for the reduce task's pair range. The reduce function generates all pairs $(x, y)$ with entity indexes $x < y$, checks if the pair index falls into the

## VI. EVALUATION

In the following we evaluate our BlockSplit and PairRange strategies regarding three performance-critical factors: the degree of data skew (Section VI-A), the number of configured map ($m$) and reduce ($r$) tasks (Section VI-B), and the number of available nodes ($n$) in the cloud environment (Section VI-C). In each experiment we examine a reasonable range of values for one of the three factors while holding constant the other two factors. We thereby broadly evaluate our algorithms and investigate to what degree they are robust against data skew, can benefit from many reduce tasks, and can scale with the number of nodes.

We ran our experiments on Amazon EC2 cloud infrastructure using Hadoop with up to 100 *High-CPU Medium* instances each providing 2 virtual cores. Each node was configured to run at most two map and reduce tasks in parallel. On each node we set up Hadoop 0.20.2 and made the same changes to the Hadoop default configuration as in [19].

We utilized two real-world datasets (see Figure 8). The first dataset DS1 contains about 114,000 product descriptions. The second dataset, DS2[4], is by an order of magnitude larger and contains about 1.4 million publication records. For both datasets, the first three letters of the product or publication title, respectively, form the default blocking key (in the robustness experiment, we vary the blocking to study skew effects). The resulting number of blocks as well as the relative size of the respective largest block are given in Figure 8. Note that the blocking attributes were not chosen to artificially generate data skew but rather reflect a reasonable way to group together similar entities. Two entities were compared by computing the edit distance of their title. Two entities with a minimal similarity of 0.8 were regarded as matches.

| Dataset | #Entities | #Blocks | #Pairs | Largest Block | |
|---------|-----------|---------|--------|---------------|---|
| | | | | #Entities | #Pairs |
| DS1 | $1.1 \cdot 10^5$ | 1,483 | $3 \cdot 10^6$ | 18% | 71% |
| DS2 | $13.9 \cdot 10^5$ | 14,659 | $6.7 \cdot 10^9$ | 4% | 26% |

Fig. 8. Datasets used for evaluation

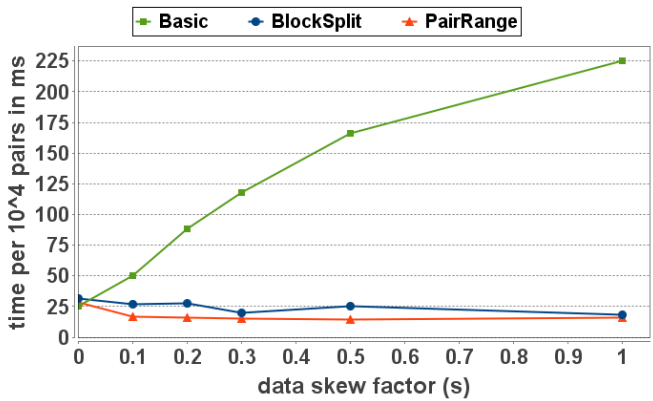

Fig. 9. Execution times for different data skews.

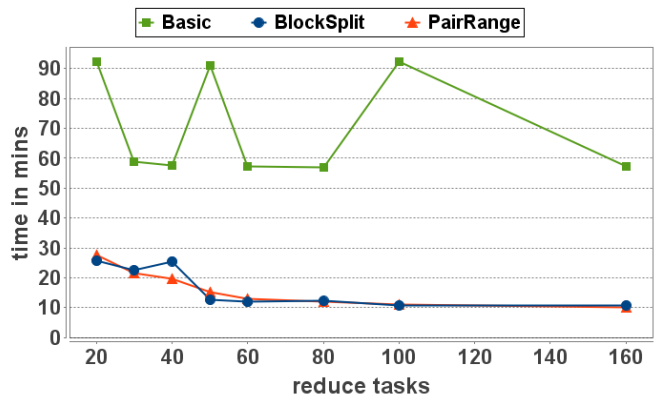

Fig. 10. Execution times for Basic, BlockSplit, and PairRange strategy using DS1.

## A. Robustness: Degree of data skew

We first evaluate the robustness of our load balancing strategies against data skew. To this end, we control the degree of data skew by modifying the blocking function and generating block distributions that follow an exponential distribution. Given a fixed number of blocks $b=100$, the number of entities in the $k^{th}$ block is proportional to $e^{-s \cdot k}$. The skew factor $s \geq 0$ thereby describes the degree of data skew. Note that the data skew, i.e., the distribution of entities over all blocks, determines the overall number of entity pairs. For example, two blocks with 25 entities each lead to $2 \cdot 25 \cdot 24/2 = 600$ pairs. If the 50 entities are split 45 vs. 5 the number of pairs equals already $45 \cdot 44/2 + 5 \cdot 4/2 = 1,000$. We are therefore interested in the average execution time *per entity pair* when comparing load balancing strategies for different data skews.

Figure 9 shows the average execution time per $10^4$ pairs for different data skews of DS1 ($n = 10$, $m = 20$, $r = 100$). The Basic strategy explained in Section III is not robust against data skew because a higher data skew increases the number of pairs of the largest block. For example, for $s=1$ Basic needs 225 ms per $10^4$ comparisons which is more than 12 times slower than BlockSplit and PairRange. However, the Basic strategy is the fastest for a uniform block distribution ($s=0$) because it does not suffer from the additional BDM computation and load balancing overhead. The BDM influence becomes insignificant for higher data skews because the data skew does not affect the time for BDM computation but the number of pairs. This is why the execution time per pair is reduced for increasing $s$. In general, both BlockSplit and PairRange are stable across all data skews with a small advantage for PairRange due to its somewhat more uniform

[4]http://asterix.ics.uci.edu/data/csx.raw.txt.gz

workload distribution.

## B. Number of reduce tasks

In our next experiment, we study the influence of the number $r$ of reduce tasks in a fixed cloud environment of 10 nodes. We vary $r$ from 20 to 160 but let the number of map tasks constant ($m = 20$). The resulting execution times for DS1 are shown in Figure 10. Execution times from PairRange and BlockSplit include the relatively small overhead (35s) for BDM computation.

We observe that both BlockSplit and PairRange significantly outperform the Basic strategy. For example, for $r=160$ they improve execution times by a factor of 6 compared to Basic.

Obviously, the Basic approach fails to efficiently leverage many reduce tasks because of its inability to distribute the matching of large blocks to multiple reduce tasks. Consequently, the required time to process the largest block (that accounts for more than 70% of all pairs, see Figure 8) forms a lower boundary of the overall execution time. Since the partitioning is done without consideration of the block size, an increasing number of reduce tasks may even increase the execution time if two or more large blocks are assigned to the same reduce task as can be seen by the peaks in Figure 10.

On the other hand, both BlockSplit and PairRange take advantage of an increasing number of reduce tasks. BlockSplit provides relatively stable execution times over the entire range of reduce tasks underlining its load balancing effectiveness. PairRange gains more from a larger number of reduce tasks and eventually outperforms BlockSplit by 7%.

However, even though PairRange always generates a uniform workload for all reduce tasks, it may be slower than BlockSplit for small $r$. This is due to the fact that the execution time is also influenced by other effects. Firstly, the execution time of a reduce task may differ due to heterogeneous hardware and matching attribute values of different length. This computational skew diminishes for larger $r$ values because of a smaller number of pairs per reduce task. Secondly, slightly unbalanced reduce task workloads can be counterbalanced by a favorable mapping to processes.

| Algorithm | Dataset | Execution Time in mins |
|---|---|---|
| BlockSplit | DS1 | 10.30 |
| | DS1 (sorted) | 18.30 |
| PairRange | DS1 | 10.38 |
| | DS1 (sorted) | 11.78 |

Fig. 11. Execution times for BlockSplit and PairRange using DS1 (unsorted/sorted by blocking key).

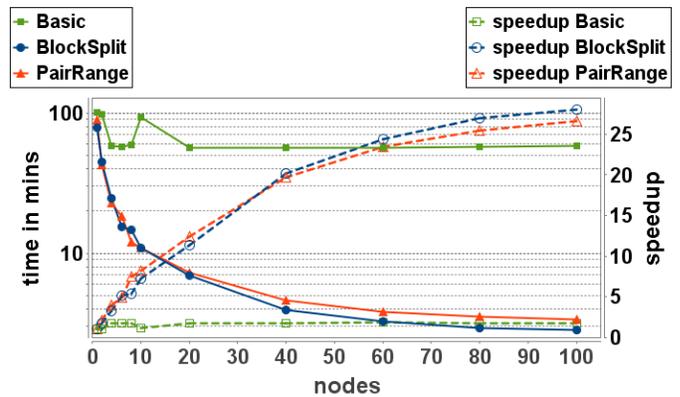

Fig. 13. Execution times and speedup of the Basic, BlockSplit, and PairRange strategy for DS1.

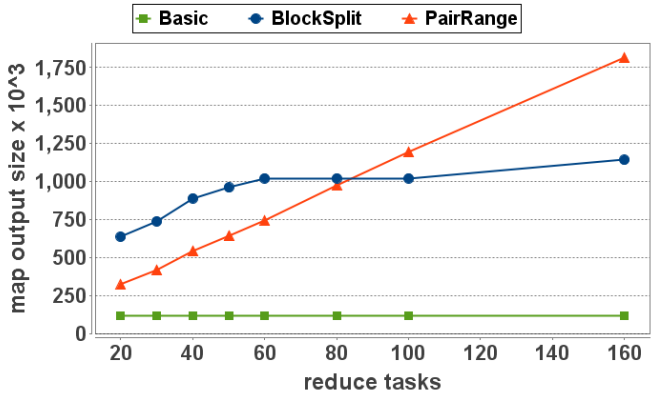

Fig. 12. Number of generated key-value pairs by map for DS1.

As we have shown in Section VI-A both strategies are *not* vulnerable to data skew but BlockSplit's load balancing strategy depends on the input (map) partitioning. To this end we have sorted DS1 by title and Figure 11 compares the execution times of for the unsorted (i.e., arbitrary order) and sorted dataset. Since the blocking key is the first three letters of the title, a sorted input dataset is likely to group together large blocks into the same map partition. This limits BlockSplit's ability to split large blocks and deteriorates its execution time by 80%. This effect can be diminished by a higher number of map tasks.

Figure 12 shows the number of emitted key-value pairs during the map phase for all strategies. The map output for Basic always equals the number of input entities because Basic does not send an entity to more than one task and, thus, does not replicate any input data. The BlockSplit strategy shows a step-function-like behavior because the number of reduce tasks determines what blocks will be split but do not influence the split method itself which is solely based on the input partitions. As a consequence, BlockSplit generates the largest map output for a small number of reduce tasks. However, an increasing number of reduce tasks increases the map output only to limited extent because large blocks that have already been split are not affected by additional reduce tasks. In contrast, the PairRange strategy is independent from the blocks but only considers pair ranges. Even though the number of relevant entities per pair range may vary (see, e.g., Figure 7) the overall number of emitted key-value pairs increases almost linearly with increasing number of ranges/ reduce tasks. For a large number of reduce tasks PairRange therefore produces the largest map output. The associated overhead (additional data transfer, sorting larger partitions) did not significantly impact the execution times up to a moderate size of the utilized cloud infrastructure due to the fact that the matching in the reduce phase is by far the dominant factor of the overall runtime. We will investigate the scalability of our strategies for large cloud infrastructures in our last experiment.

### C. Scalability: Number of nodes

Scalability in the cloud is not only important for fast computation but also for financial reasons. The number of nodes should be carefully selected because cloud infrastructure vendors usually charge per employed machines even if they are underutilized. To analyze the scalability of Basic, BlockSplit, and PairRange, we vary the number of nodes from 1 up to 100. For $n$ nodes, the number of map tasks is set to $m = 2 \cdot n$ and the number of reduce tasks is set to $r = 10 \cdot n$, i.e., adding new nodes leads to additional map and reduce tasks. The resulting execution times and speedup values are shown in Figure 13 (DS1) and Figure 14 (DS2).

As expected, Basic does not scale for more than two nodes due to the limitation that all entities of a block are compared within a single reduce task. The execution time is therefore dominated by the reduce task that has to process the largest block and, thus, about 70% of all pairs. An increasing number of nodes only slightly decreases the execution time because the increasing number of reduce tasks reduces the additional workload of the reduce task that handles the largest block.

By contrast, both BlockSplit and PairRange show their ability to evenly distribute the workload across reduce tasks and nodes. They scale almost linearly up to 10 nodes for the smaller dataset DS1 and up to 40 nodes for the larger dataset DS2, respectively. For large $n$ we observe significantly better speedup values for DS2 than for DS1 due to the reasonable workload per reduce task that is crucial for efficient utilization of available cores. BlockSplit outperforms PairRange for DS1 and $n$=100 nodes. The resulting large number of reduce tasks leads – in conjunction with the comparatively small data size – to a comparatively small average number of comparisons per reduce task. Therefore PairRange's additional overhead (see Figure 12) deteriorates the overall execution time. This overhead becomes insignificant for the larger dataset DS2. The average number of comparisons is more than 2,000 times higher than for DS1 (see Figure 8) and, thus, the benefit of optimally balanced reduce tasks outweighs the additional overhead of handling more key-value pairs. In general, BlockSplit is preferable for smaller (splittable) datasets under the assumption that the dataset's data order is not dependent from the blocking key; otherwise PairRange has a better performance.

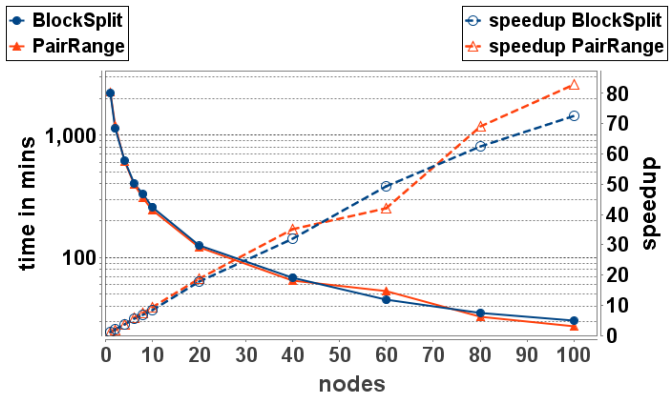

Fig. 14. Execution times and speedup of the BlockSplit and PairRange strategy for DS2.

## VII. RELATED WORK

Load balancing and skew handling are well-known data management problems and MR has been criticized for having overlooked the skew issue [7]. Parallel database systems already implement skew handling mechanisms, e.g., for parallel hash join processing [6] that share many similarities with our problem.

A theoretical analysis of skew effects for MR is given in [15] but focuses on linear processing of entities in the reduce phase. It disregards the N-squared complexity of comparing all entities with each other. [14] reports that the reduce runtime for scientific tasks does not only depend on the assigned workload (e.g., number of pairs) but also on the data to process. The authors propose a framework for automatic extraction of signatures for (spatial) data to reduce the computational skew. This approach is orthogonal to ours: it addresses computational skew and does not consider effective handling of present data skew.

A fairness-aware key partitioning approach for MR that targets locality-aware scheduling of reduce tasks is proposed in [10]. The key idea is to assign map output to reduce tasks that eventually run on nodes that already hold a major part of the corresponding data. This is achieved by a modification of the MR framework implementation to control the scheduling of reduce tasks. Similar to our BDM, this approach determines the key distribution to optimize the partitioning. However, it does not split large blocks but still processes all data sharing the same key at the same reduce task which may lead to unbalanced reduce workloads.

MR has already been employed for ER (e.g., [20]) but we are only aware of one load balancing mechanism for MR-based ER. [11] studies load balancing for Sorted Neighborhood (SN). However, SN follows a different blocking approach that is by design less vulnerable to skewed data.

MR's inherent vulnerability to data skew is relevant for all kind of pairwise similarity computation. Example applications include pairwise document similarity [9] to identify similar documents, set-similarity joins [19] for efficient string similarity computation in databases, pairwise distance computation [18] for clustering complex objects, and all-pairs matrix computation [16] for scientific computing. All approaches follow a similar idea like ER using blocking: One or more signatures (e.g., tokens or terms) are generated per object (e.g., document) to avoid the computation of the Cartesian product. MR groups together objects sharing (at least) one signature and performs similarity computation within the reduce phase. Simple approaches like [9] create many signatures per object which leads to unnecessary computation because similar objects are likely to have more than one signature in common and are thus compared multiple times. Advanced approaches such as [19] reduce unnecessary computation by employing filters (e.g., based on token frequencies) that still guarantee that similar object pairs share at least one signature.

A more general case is the computation of theta-joins with MapReduce [17]. Static load balancing mechanisms are not suitable due to arbitray join conditions. Similar to our approach [17] employs a pre-analysis phase to determine the datasets' characteristics (using sampling) and thereby avoids the evaluation of the Cartesian product. This approach is more coarse-grained when compared to our strategies.

Load balancing is only one aspect towards an optimal execution of MR programs. For example, Manimal [4] employs static code analysis to optimize MR programs. Hadoop++ [8] proposes index and join techniques that are realized using appropriate partitioning and grouping functions, amongst others.

## VIII. SUMMARY AND OUTLOOK

We proposed two load balancing approaches, BlockSplit and PairRange, for parallelizing blocking-based entity resolution using the widely available MapReduce framework. Both approaches are capable to deal with skewed data (blocking key) distributions and effectively distribute the workload among all reduce tasks by splitting large blocks. Our evaluation in a real cloud environment using real-world data demonstrated that both approaches are robust against data skew and scale with the number of available nodes. The BlockSplit approach is conceptionally simpler than PairRange but achieves already excellent results. PairRange is less dependent on the initial partitioning of the input data and slightly more scalable for large match tasks.

In future work, we will extend our approaches to multi-pass blocking that assigns multiple blocks per entity. We will further investigate how our load balancing approaches can be adapted for MapReduce-based implementations of other data-intensive tasks, such as join processing or data mining.

# APPENDIX I
## MATCHING TWO SOURCES

This section describes the extension of the BlockSplit and PairRange strategy for matching two sources $R$ and $S$. We thereby assume that all entities have a valid blocking key. Consideration of entities without valid blocking keys can be accomplished as follows:

$$\text{match}_B(R, S) = \text{match}_B(R - R_\emptyset, S - S_\emptyset)$$
$$\cup \text{match}_\perp(R, S_\emptyset) \cup \text{match}_\perp(R_\emptyset, S - S_\emptyset)$$

Given two sources $R$ an $S$ with a subset $R_\emptyset \subseteq R$ and $S_\emptyset \subseteq S$ of entities without blocking keys, the desired match result $\text{match}_B(R, S)$ using a blocking key $B$ can be constructed as union of three match results. First, the regular matching is applied for entities with valid blocking keys only ($R - R_\emptyset$ and $S - S_\emptyset$). The result is then completed with the match results of the Cartesian product of $R$ with $S_\emptyset$ and $R_\emptyset$ with $S - S_\emptyset$. Such results can be obtained by employing a constant blocking key (denoted as $\perp$) so that all entity pairs are considered.

For simplicity we furthermore assume that each input partition contains only entities of one source (this can be ensured

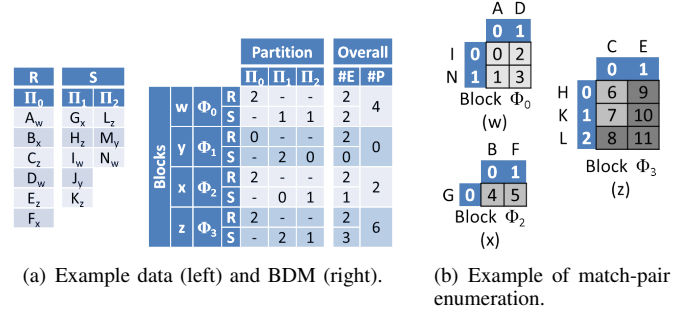

(a) Example data (left) and BDM (right).  (b) Example of match-pair enumeration.

Fig. 15. Matching two sources R and S.

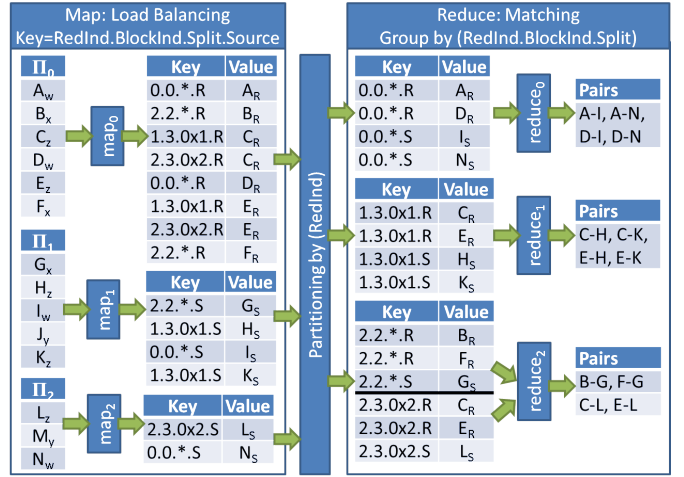

Fig. 16. Example BlockSplit dataflow for 2 sources.

by Hadoop's *MultipleInputs* feature). The number of partitions may be different for each of the two sources.

For illustration, we use the example data of Figure 15(a) that utilizes the entities $A$-$N$ and the blocking keys $w$-$z$. Each entity belongs to one of the two sources $R$ and $S$. Source $R$ is stored in one partition $\Pi_0$ only whereas entities of $S$ are distributed among two partitions $\Pi_1$ and $\Pi_2$.

The BDM computation is the same but adds a source tag to the map output key to identify blocks with the same key in different sources, i.e., $\Phi_{i,R}$ and $\Phi_{i,S}$. The BDM has the same structure as for the one-source case but distinguishes between the two sources for each block (see Figure 15(a)).

### A. Block-based Load Balancing

The BlockSplit strategy for two sources follows the same scheme as for one source. The main difference is that the keys are enriched with the entities' source and that each entity (value) is annotated with its source during the map phase. Hence, map outputs key-value pairs with key=(reduce_index $\odot$ block_index $\odot$ split $\odot$ source) and value=(entity). This allows the reduce phase to easily identify all pairs of entities from different sources. Like in the one-source case BlockSplit splits large blocks $\Phi_k$ but restricts the resulting match tasks $k.i \times j$ so that $\Pi_i \in R$ and $\Pi_j \in S$.

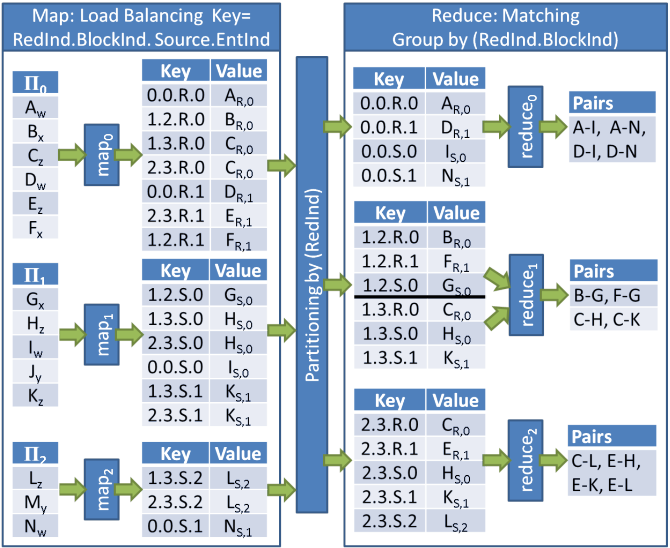

Fig. 17. Example PairRange dataflow for 2 sources.

Figure 16 shows the workflow for the example data of Figure 15(a). The BDM indicates 12 overall pairs so that the average reduce workload equals 4 pairs. The largest block $\Phi_3$ is therefore subject to split because it has to process 6 pairs. The split results in the two match tasks $3.0\times1$ and $3.0\times2$. All match tasks are ordered by the number of pairs: $0.*$ (4 pairs, reduce$_0$), $3.0\times1$ (4 pairs, reduce$_1$), $2.*$ (2 pairs, reduce$_2$), $3.0\times2$ (2 pairs, reduce$_2$). The eventual dataflow is shown in Figure 16. Partitioning is based on the reduce task index only, for routing all data to the reduce tasks whereas sorting is done based on the entire key. The reduce function is called for every match task $k.i \times j$ and compares entities considering only pairs from different sources. Thereby, the reduce tasks read all entities of $R$ and compare each entity of $S$ to all entities of $R$.

### B. Pair-based Load Balancing

The PairRange strategy for two sources follows the same principles as for one source. Entity enumeration is realized per block and source like in Section V. Pair enumeration is done for blocks $\Phi_i$ using entities of $\Phi_{i,R}$ and $\Phi_{i,S}$ sharing the same blocking key. The enumeration scheme is column-oriented but all cells of the $|\Phi_{i,R}| \times |\Phi_{i,S}|$ matrix will be enumerated. For two entities $e_R \in \Phi_{i,R}$ and $e_S \in \Phi_{i,S}$ with entity indexes $x$ and $y$, respectively, the pair index is defined as follows: $p_i(x,y) = c(x,y,|\Phi_{i,S}|) + o(i)$ with $c(x,y,N) = x \cdot N + y$ and $o(i) = \Sigma_{k=0}^{i-1}(|\Phi_{k,R}| \cdot |\Phi_{k,S}|) - 1$. Figure 15(b) shows the resulting match-pair enumeration for our running example. With $r = 3$, the resulting 12 pairs are divided into three ranges of size 4. Block $\Phi_1$ (blocking key equals y) needs not to be considered because no entity in source $S$ has such a blocking key.

The map phase identifies all relevant ranges for each entity. For an entity $e_R \in \Phi_{i,R}$ with index $x$ the ranges of pairs $p_i(x,0)$ through $p_i(x,|\Phi_{i,S}|)$ need to be considered whereas for $e_S \in \Phi_{i,S}$ with index $y$ the pairs $p_i(0,y)$ through $p_i(|\Phi_{i,R}|,y)$ are relevant.

For each relevant range, map emits a key-value pair with key=(range_index ⊙ block_index ⊙ source ⊙ entity_index) and value= entity. Compared to the the one-source case, in addition to its entity index, each entity (value) is also annotated with its source ($R$ or $S$). Partitioning is based on the range index only, for routing all data to the reduce tasks. Sorting is done based on the entire key. The reduce function is called for every block and compares entities like in the one-source case but only considers pairs from different sources.

Figure 17 illustrates the approach using the example data of Figure 15(a). For example, entity $C \in R$ is the first entity (index=0) within block $\Phi_3$. It takes part in ranges $\Re_1$ and $\Re_2$ and will therefore be sent to the second and third reduce task. Hence map emits two keys $(1.3.R.0)$ and $(2.3.R.0)$, respectively, for entity $C$.

## APPENDIX II
## LISTINGS

In the following, we show the pseudo-code for the two proposed load balancing strategies and the BDM computation. Beside the regular output of Algorithm 3 (the BDM itself), map uses a function *additionalOutput* that writes each entity along with its computed blocking key to the distributed file system. The additional output of the first MR job is read by the second MR job. By prohibiting the splitting of input files, it is ensured that the second MR job receives the same partitioning of the input data as the first job. A map task of the second job processes exactly one additional output file (produced by a map task of the first task) and can extract the corresponding partition index from the file name. With the help of Hadoop's data locality for map task assignment, it is likely that there is *no* redistribution of additional output data.

The map tasks of the second job read the BDM at initialization time. It is not required that each map task holds the full BDM in memory. For each blocking key that occurs in the respective map input partition, it is sufficient to store the overall sum of entities in previous map input partitions (Algorithm 2 Lines 4-8). Furthermore, it would be possible to store the BDM in a distributed storage like HBase to avoid memory shortcomings.

For readability, the pseudo-code refers to the following functions:

- **BDM.blockIndex(blockKey)** returns the block's index
- **BDM.size(blockIndex)** returns #entities for a given block
- **BDM.size(blockIndex, partitionIndex)** returns #entities for a given block in this partition
- **BDM.pairs()** returns overall number of entity pairs
- **getNextReduceTask** returns the reduce task with the fewest number of assigned entity comparisons (BlockSplit)
- **addCompsToReduceTask(reduceTask, comparisons)** increases number of assigned pairs of the given reduce task by the given value (BlockSplit)
- **match(e$_1$, e$_2$)** compares two entities and adds matching pairs to the final output.

## Algorithm 1: Implementation of BlockSplit

```
1  map_configure (m, r, partitionIndex)
2      matchTasks ← empty map;
3      compsPerReduceTask ← BDM.pairs()/r;
4      // Read BDM from reduce output of Algorithm 3
5      BDM ← readBDM();
6      // Match task creation
7      for k ← 0 to BDM.numBlocks()-1 do
8          comps ← ½ · BDM.size(k) · (BDM.size(k) − 1);
9          if comps ≤ compsPerReduceTask then
10             matchTasks.put((k, 0, 0), comps);
11         else
12             for i ← 0 to m-1 do
13                 |Φ_k^i| ← BDM.size(k, i);
14                 for j ← 0 to i do
15                     |Φ_k^j| ← BDM.size(k, j);
16                     if |Φ_k^i| · |Φ_k^j| > 0 then
17                         if i = j then
18                             matchTasks.put((k, i, j),
19                                 ½ · |Φ_k^i| · (|Φ_k^i| − 1));
20                         else
21                             matchTasks.put((k, i, j), |Φ_k^i| · |Φ_k^j|);

22     // Reduce task assignment
23     matchTasks.orderByValueDescending();
24     foreach ((k,i,j), comps) ∈ matchTasks do
25         reduceTask ← getNextReduceTask();
26         matchTasks.put((k, i, j), reduceTask);
27         addCompsToReduceTask(reduceTask, comps);

28 // Operate on additional map output of Algorithm 3
29 map (k_in=blockingKey, v_in=entity)
30     k ← BDM.blockIndex(blockingKey);
31     comps ← ½ · BDM.size(k) · (BDM.size(k) − 1);
32     if comps ≤ compsPerReduceTask then
33         if comps>0 then
34             reduceTask ← matchTasks.get(k, 0, 0);
35             output(k_tmp=reduceTask.k.0.0,
36                 v_tmp=(entitiy, partitionIndex));
37     else
38         for i ← 0 to m-1 do
39             min ← min(partitionIndex, i);
40             max ← max(partitionIndex, i);
41             reduceTask ← matchTasks.get(k, max, min);
42             if reduceTask ≠ null then
43                 output(k_tmp=reduceTask.k.max.min,
44                     v_tmp=(entitiy, partitionIndex));

45 // part: Repartition map output by reduceTask
46 // comp: Sort by blockIndex.i.j (k.i.j)
47 // group: Group by blockIndex.i.j (k.i.j)
48 reduce (k_tmp=reduceTask.k.i.j,
49     list(v_tmp)=list((entity, partitionIndex)))
50     buffer ← [];
51     if i = j then
52         foreach (e_2, partitionIndex) ∈ list(v_tmp) do
53             foreach e_1 ∈ buffer do
54                 match(e_1, e_2);
55             buffer.append(e_2);
56     else
57         pair ← list(v_tmp).firstElement();
58         buffer.append(pair.first);
59         firstPartitionIndex ← pair.second();
60         foreach (e_2, partitionIndex) ∈ list(v_tmp) do
61             if partitionIndex = firstPartitionIndex then
62                 buffer.append(e_2);
63             else
64                 foreach e_1 ∈ buffer do
65                     match(e_1, e_2);
```

## Algorithm 2: Implementation of PairRange

```
1  map_configure (m, r, partitionIndex)
2      BDM ← readBDM() ;           // Output of MR job 1
3      compsPerReduceTask ← ⌈BDM.pairs()/r⌉;
4      entityIndex ← [] ;    // Next entity index for each block
5      for i ← 0 to BDM.numBlocks()-1 do
6          entityIndex[i] ← 0;
7          for j ← 0 to partitionIndex-1 do
8              entityIndex[i] ← entityIndex[i]+ BDM.size(i, j)

9  // Operate on additional map output of Algorithm 3
10 map (k_in=blockingKey, v_in=entity)
11     ranges ← ∅;
12     i ← BDM.blockIndex(blockingKey);
13     x ← entityIndex[i];
14     N ← BDM.size(i);
15     ℜ_min ← rangeIndex(0, max(x, 1), N, i);
16     ℜ_max ← rangeIndex(min(x, N-2), N-1, N, i);
17     ranges ← {ℜ_min} ∪ {ℜ_max};
18     if ranges.size>2 then
19         for k ← 1 to x-1 do
20             ranges ← ranges ∪{k};
21         ℜ_med ← rangeIndex(min(x, N-2), min(x+1, N-1), N, i);
22         for k ← ℜ_med to ℜ_max-1 do
23             ranges ← ranges ∪ {k};

24     foreach r ∈ ranges do
25         output (k_tmp=r.i.x, v_tmp=(entitiy, x));

26     entityIndex[i] ← entityIndex[i]+1;

27 reduce_configure (m, r)
28     BDM ← readBDM();
29     compsPerReduceTask ← ⌈BDM.pairs()/r⌉;

30 // Repartition map output by range index (r), sort by
31 // blockIndex.entityIndex (i.x), group by blockIndex (i)
32 reduce (k_tmp=r.i.x, list(v_tmp)=list((entity, x)))
33     N ← BDM.size(i);
34     buffer ← [];
35     foreach (e_2, x_2) ∈ list(v_tmp) do
36         foreach (e_1, x_1) ∈ buffer do
37             k ← rangeIndex (x_1, x_2, N, i);
38             if k=r then
39                 match(e_1, e_2) ; // Comparison; output matches
40             else if k>r then
41                 return;

42         buffer.append((e_2, x_2));

43 rangeIndex (col, row, blockSize, blockIndex)
44     cellIndex ← 0.5 · col · (2·blockSize-col-3)+row-1;
45     pairIndex ← cellIndex + pairIndexOffset(blockIndex);
46     return ⌊pairIndex/compsPerReduceTask⌋;

47 pairIndexOffset (blockIndex)
48     sum ← 0;
49     for k ← 0 to blockIndex-1 do
50         sum ← BDM.size(k)·(BDM.size(k)-1) + sum;
51     return 0.5 · sum;
```

## Algorithm 3: Computation of the BDM

```
1  map_configure (m, r, partitionIndex)
2      // Store partitionIndex

3  map (k_in=unused, v_in=entity)
4      blockingKey = computeKey(entity);
5      additionalOutput (k=blockingKey, v=entity) ;      // to DFS
6      output (k_tmp =blockingKey.partitionIndex, v_tmp=1);

7  // Repartition map output by blockingKey, sort by
8  // blockingKey.partitionIndex, group by blockingKey.partitionIndex
9  reduce (k_tmp=blockingKey.partitionIndex, list(v_tmp)=list(1)))
10     sum ← 0;
11     foreach number in list(v_tmp) do
12         sum ← sum+number;

13     out ← blockingKey +","+partitionIndex +","+sum;
14     output (k_out=unused, v_out=out);
```